\documentclass[aip,sd,amsmath,amssymb,reprint]{revtex4-1}

\usepackage{graphicx}
\usepackage{dcolumn}
\usepackage{bm}
\usepackage{epstopdf}
\usepackage{xcolor}
\usepackage{natbib}
\usepackage{subfigure}

\newcommand{\rc}[1]{\textcolor{black}{#1}}

\begin{document}

\title{Birefringence in thermally anisotropic relativistic plasmas and its impact on laser-plasma interactions}


\author{A. Arefiev}
\affiliation{Department of Mechanical and Aerospace Engineering, University of California at San Diego,\\La Jolla,
CA 92093, USA}

\author{D. J. Stark}
\affiliation{Los Alamos National Laboratory, Los Alamos, New Mexico 87545, USA}

\author{T. Toncian}
\affiliation{Institute for Radiation Physics, Helmholtz-Zentrum Dresden-Rossendorf e.V., 01328 Dresden, Germany}

\author{M. Murakami}
\affiliation{Institute of Laser Engineering, Osaka University, Osaka 565-0871, Japan}

\date{\today}

\begin{abstract}
{One of the paradigm-shifting phenomena triggered in laser-plasma interactions at relativistic intensities is the so-called relativistic transparency. As the electrons become heated by the laser to relativistic energies, the plasma becomes transparent to the laser light even though the plasma density is sufficiently high to reflect the laser pulse in the non-relativistic case. This paper highlights the impact that relativistic transparency can have on laser-matter interactions by focusing on a collective phenomenon that is associated with the onset of relativistic transparency: plasma birefringence in thermally anisotropic relativistic plasmas. The optical properties of such a system become dependent on the polarization of light, and this can serve as the basis for plasma-based optical devices or novel diagnostic capabilities.}
\end{abstract}

\maketitle


\section{Introduction}

\rc{The advent of ultra-high intensity lasers has precipitated a corresponding effort to better characterize the fundamental physics at play in high-amplitude laser-matter interactions.} A laser pulse of high intensity irradiating a solid material quickly turns it into a plasma and then continues to heat the plasma electrons. As the electron motion becomes relativistic due to the heating induced by the laser pulse, the very optical properties of the plasma change. The plasma can become transparent to the laser light even though its density is sufficiently high to reflect the laser pulse in the case of non-relativistic electron energies~\cite{kaw1970relativistic,max1971strong,eremin2010relativistic,cattani2000threshold,tushentsov2001electromagnetic,goloviznin2000self,siminos2012effect,weng2012ultra,bulanov2013strong,vshivkov1998nonlinear}. This fundamentally alters the nature of the interaction and has far-reaching repercussions on the subsequent evolution of the laser-plasma system. 

The goal of this paper is to highlight a novel collective phenomenon that is associated with the onset of relativistic transparency: plasma birefringence in anisotropic momentum distributions~\cite{Stark-PhysRevLett.115.025002}. Here the optical properties of a thermally anisotropic plasma become dependent on the polarization of the laser light. An accurate description of the polarization dependent dispersion in relativistic plasmas finds its application in the generation of plasmonic devices \cite{Gonzalez2016, Lehmann2016,lehmann2018plasma, Turnbull2017}, which have garnered much attention due to their tunability and damage-resistant nature. The threshold for relativistic transparency in laser-plasma interactions has long been a topic of study~\cite{siminos2012effect,cattani2000threshold,goloviznin2000self,eremin2010relativistic,siminos2017kinetic,gelfer2020absorption} because of its impact on the nature of energy transfer and particle dynamics; in particular, ion acceleration studies often operate at densities in the relativistic transparency regime~\cite{palaniyappan2012dynamics,yin2011break,fernandez2017laser,stark2018harnessing,stark2018detailed} and rely on precise characterizations of this threshold \cite{Bychenkov2016}. Understanding the role of polarization on these thresholds is critical for understanding the underlying physics at play in these systems, and it can also potentially serve as the basis for developing distribution function diagnostics in laboratory studies.

First experimental validations of polarization rotation due to anisotropy of plasma heated by a linearly polarized relativistic intensity laser pulse can be found in Refs.~[\onlinecite{Gonzalez2016}] and~[\onlinecite{duff2020high}], and here we delineate in detail the origins and applications of this effect. Section~\ref{Sec-RT} gives the analytical treatment of the electromagnetic dispersion relation in a relativistic plasma with momentum anisotropy, Section~\ref{Sec-MT} provides particle-in-cell (PIC) simulations demonstrating the birefrigent properties of these plasmas, and we close with a discussion in Section~\ref{Sec-C}.


\section{Analytical treatment of the optical properties of relativistic plasmas} \label{Sec-RT}

Electron heating by an irradiating laser pulse can fundamentally alter the optical properties of the plasma. In this section, we show that not only an otherwise opaque plasma can become transparent to an electromagnetic wave due to relativistic electron motion, but it can also become birefringent. While the effect of relativistically induced transparency is well-known, the relativistically induced birefringence, where the optical properties of the medium become dependent on the polarization of light, is a newly discovered effect~\cite{Stark-PhysRevLett.115.025002}. 

We begin by reviewing key properties of light propagation through a warm classical non-relativistic plasma. In the case of a plasma with an electron density $n_e$ and an electron temperature $T_e$, the dispersion relation for a linear electromagnetic wave is given by~\cite{Ginzburg1964}:
\begin{equation} \label{disp-relat}
\omega^2 = \omega_{pe}^2 + c^2 \left( 1+ \frac{T_e}{m_e c^2} \frac{\omega_{pe}^2}{\omega^2} \right) k^2,
\end{equation}
in which $\omega$ is the wave frequency, $k$ is its wave vector, and 
\begin{equation}
\omega_{pe} = \sqrt{4 \pi n_e e^2/ m_e}
\end{equation}  
is the plasma frequency. Here $c$ is the speed of light and $m_e$ and $e$ are the electron mass and charge respectively. It follows directly from Eq.~(\ref{disp-relat}) that the electromagnetic wave can propagate only if the electron density is below the classical critical density defined as
\begin{equation} \label{crit-classical}
n_* \equiv \frac{m_e \omega^2}{4 \pi e^2}.
\end{equation}
The critical density is independent of the electron temperature, meaning that electron heating is inconsequential when it comes to transparency of a non-relativistic plasma. 

However, if the plasma electrons are heated to relativistic energies, then the relation between the electron momentum $p$ and velocity $v$ changes from  $p = m_e v$ to $p = \gamma m_e v$, where $\gamma = \sqrt{1 + p^2/m_e^2 c^2}$ is the relativistic factor. This  can effectively be interpreted as an increase in the electron mass. One can then expect, based on Eq.~(\ref{crit-classical}), that the critical density would increase by a factor of $\gamma$ as well compared to the classical critical density $n_*$:
\begin{equation} \label{RT}
n_{crit} \approx \gamma n_*.
\end{equation}
This means that a plasma with an otherwise opaque electron density, $n_* < n_e <  n_{crit} \approx \gamma n_*$, would become transparent to an electromagnetic wave if the electrons are heated to relativistic energies $\sim \gamma m_e c^2$. The effect is often referred to as relativistic transparency.

\begin{figure}[htb]
	\centering
	\includegraphics[width=0.8\columnwidth]{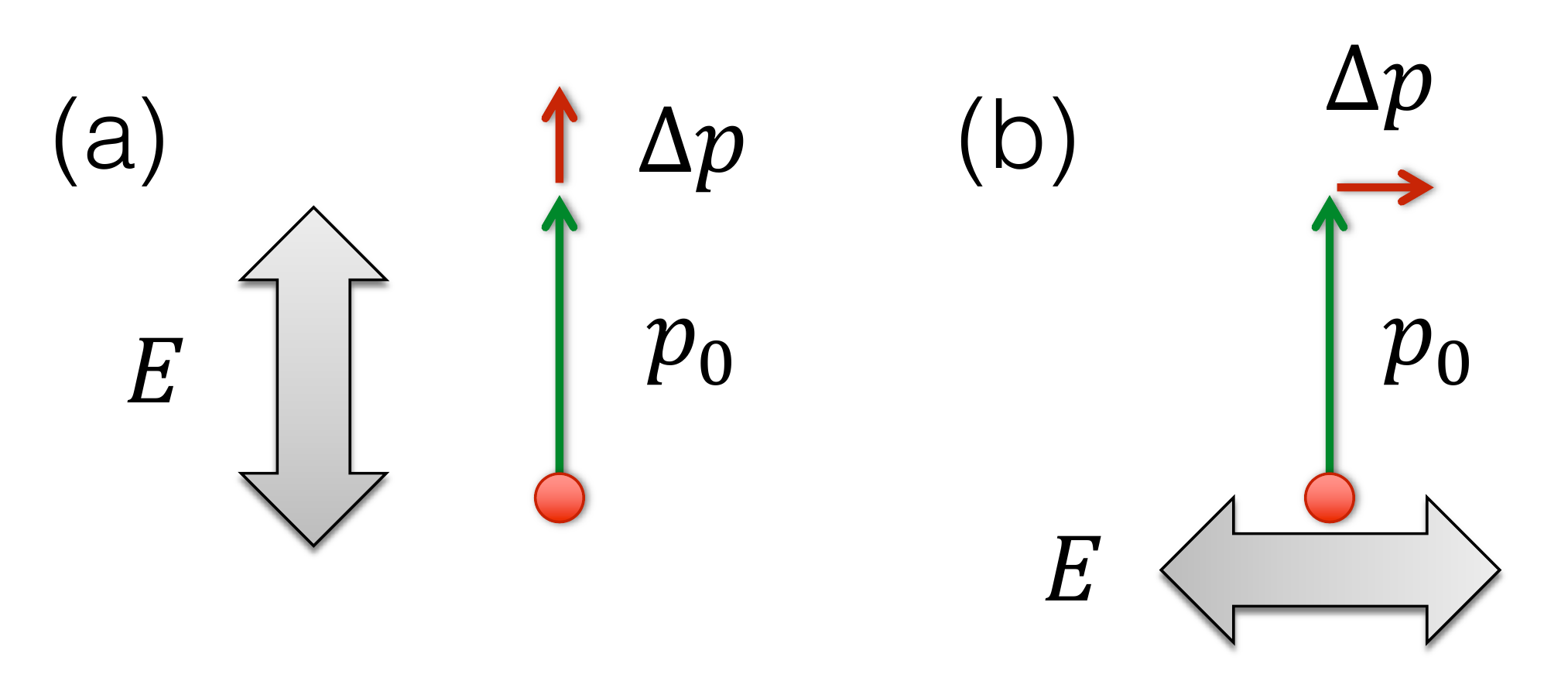} 
  \caption{Two cases of the laser electric field $E$ collinear (a) and orthogonal (b) to the electron momentum $p_0$.} \label{Figure_RT_1}
\end{figure}

This transparency is caused by the reduction in the electron current, because ``heavier'' electrons are less efficient in generating current in response to an electromagnetic wave. While the presented argument captures a well-known effect, it overlooks how the electron current is actually driven by the wave. In order to illustrate an important subtlety, let us consider how the velocity of a relativistic electron changes in response to a low-amplitude electromagnetic wave. The two cases of interest are shown in Fig.~\ref{Figure_RT_1}. If the electric field is collinear to the electron momentum $p_0$, then a change of the momentum by $\Delta p$ leads to a change in the electron velocity that can be estimated as
\begin{equation}
\frac{\Delta v}{c} \approx \frac{1}{\gamma^2} \frac{\Delta p}{p_0},
\end{equation}
where $\gamma = \sqrt{1 + p_0^2 / m_e^2 c^2}$.
On the other hand, if the electric field is orthogonal to the electron momentum $p_0$, then a change of the momentum by $\Delta p$ leads to a change in the electron velocity by
\begin{equation}
\frac{\Delta v}{c} \approx \frac{\Delta p}{p_0}.
\end{equation}
Evidently, the velocity change is greatly reduced if the electric field is collinear with the momentum of the relativistic electron. This is because the velocity of the electron is already close to the speed of light in the direction of the intended change, so any potential change can only be relatively small. 

These estimates indicate that the electron current should have a strong dependence on the relative orientation of \rc{laser electric field} $\textbf{E}$ and \rc{particle momentum} $\textbf{p}_0$. However, if the electron momentum distribution is isotropic, then the anisotropic effect from individual electrons would be negated in the total electron current. On the other hand, an anisotropic distribution should enable this effect to manifest itself on the macroscopic level as a birefringent response to an electromagnetic wave, as the electrons would have a preferred direction to their momentum vectors. 

In order to quantitatively examine optical properties of an anisotropic relativistic plasma, we consider a simplified setup where the plasma is irradiated by a low-amplitude electromagnetic wave. The unperturbed plasma is assumed to be uniform and anisotropic in momentum space. Our goal is to derive a dispersion relation for the low-amplitude wave following a standard approach.

A linearized kinetic equation for the plasma electrons,
\begin{equation}
\frac{\partial f}{\partial t} + {\bf{v}} \frac{\partial f}{\partial {\bf{r}}}  - |e| \left( {\bf{E}} + \frac{1}{c} \left[ {\bf{v}} \times {\bf{B}} \right] \right) 
\frac{\partial F}{\partial {\bf{p}}} = 0,
\end{equation}
has the following form in Fourier representation: 
\begin{equation}
i (k_{\mu} v_{\mu} - \omega) f - |e| \left( E_s + \frac{1}{c} \left[ {\bf{v}} \times {\bf{B}} \right]_s \right) 
\frac{\partial F}{\partial p_s} = 0,
\end{equation}
where $f$ is the perturbation to the distribution function $F$ induced by electric and magnetic fields \textbf{E} and \textbf{B}. Here $\omega$ and \textbf{k} are the corresponding frequency and wave-vector of the perturbation and \textbf{v} and \textbf{p} are the electron velocity and momentum, respectively. Using the definition of the electron current,
\begin{equation}
j_{\alpha} \equiv - \int |e| v_{\alpha} f  d^3 p, 
\end{equation}
and taking into account that 
\begin{equation}
{\bf{B}} = \frac{c}{\omega} \left[ {\bf{k}} \times {\bf{E}} \right],
\end{equation}
one can find that
\begin{equation}
j_{\alpha} = \int \frac{i e^2 v_{\alpha} E_{\beta}}{k_{\mu} v_{\mu} - \omega} 
\left[ \delta_{s \beta} \left( 1 - \frac{k_{\mu} v_{\mu}}{\omega} \right) + \frac{k_s v_{\beta}}{\omega} \right]
\frac{\partial F}{\partial p_s}  d^3 p.
\end{equation}
A general expression for the dielectric tensor is then   
\begin{eqnarray} \label{disp_tensor_1}
\varepsilon_{\alpha \beta} &=& \delta_{\alpha \beta} + \frac{4 \pi e^2}{\omega^2}
\int \frac{\partial F}{\partial p_{\beta}} v_{\alpha} d^3 p \nonumber \\ 
&-& \frac{4 \pi e^2}{\omega^2} \int \frac{v_{\alpha} v_{\beta} k_s}{k_{\mu} v_{\mu} - \omega} \frac{\partial F}{\partial p_s}  d^3 p. 
\end{eqnarray}

In order to provide a striking example of relativistically induced birefringence, we consider a plasma that consists of two counter-streaming relativistic flows:
\begin{equation} \label{EDF_ts}
F = \frac{1}{2} \left[ n_0 \delta( {\bf{p}} - {\bf{p}}_0) + n_0 \delta( {\bf{p}} + {\bf{p}}_0) \right].
\end{equation}
where $\textbf{p}_0 = m_e \textbf{u} / \sqrt{1 - u^2/c^2}$ is the electron momentum associated with the flow velocity \textbf{u}. The corresponding dielectric tensor that follows from Eq.~(\ref{disp_tensor_1}) is 
\begin{eqnarray} \label{epsilon_2}
\varepsilon_{\alpha \beta} &=& \delta_{\alpha \beta} \left( 1 - \frac{1}{\gamma} \frac{\omega_p^2}{\omega^2} \right) \nonumber \\
&+& \frac{1}{\gamma} \frac{\omega_p^2}{\omega^2} \frac{u_{\alpha} u_{\beta}}{c^2}
\frac{\left( \omega^2 - k^2 c^2 \right) \left( \omega^2 + (k_{\mu} u_{\mu})^2 \right)}{\left( k_{\mu} u_{\mu} - \omega \right)^2 \left( k_{\mu} u_{\mu} + \omega \right)^2} \nonumber \\
&+& \frac{1}{\gamma} \frac{\omega_p^2}{\omega^2} \frac{k_{\mu} u_{\mu} \left( k_{\alpha} u_{\beta} + k_{\beta} u_{\alpha} \right)}{\left( k_{\mu} u_{\mu} - \omega \right)  \left( k_{\mu} u_{\mu} + \omega \right)}.
\end{eqnarray}
If the counter-streaming flows are non-relativistic, then a simplified expression follows from Eq.~(\ref{epsilon_2}) by setting $\gamma = 1$. 

We simplify the analysis by assuming that the wave propagation is transverse to the counter-streaming flows, such that $(\textbf{k} \cdot \textbf{u}) = 0$. Let us also assume without any loss of generality that both \textbf{k} and \textbf{u} are in the $(x,y)$-plane and that $\textbf{u} = u \textbf{e}_y$ and $\textbf{k} = k \textbf{e}_x$. The wave dispersion relations are determined from a general condition
\begin{equation} \label{dispersion_general}
\mbox{det} \left[  \frac{k_{\alpha} k_{\beta} c^2}{\omega^2} - \frac{k^2 c^2}{\omega^2} + \varepsilon_{\alpha \beta} \right] = 0 .
\end{equation}
For the dielectric tensor given by Eq.~(\ref{epsilon_2}), this equation leads to
\begin{eqnarray} \label{dispersion_counter}
&& \left[ 1 -  \frac{1}{\gamma} \frac{\omega_p^2}{\omega^2} \right] 
\left[ 1 -  \frac{1}{\gamma^3} \frac{\omega_p^2}{\omega^2} - \left(1 + \frac{1}{\gamma} \frac{\omega_p^2}{\omega^2} \frac{u^2}{c^2} \right) \frac{k^2 c^2}{\omega^2}  \right]
 \nonumber \\
&&  \left[ 1 -  \frac{1}{\gamma} \frac{\omega_p^2}{\omega^2} - \frac{k^2 c^2}{\omega^2} \right] = 0.
\end{eqnarray}
The three dispersion relations that follow are 
\begin{eqnarray} \label{dispersion_counter}
&& \omega^2 = \omega_p^2 \left/ \gamma \right., \label{CS-1} \\ 
&&  \omega^2 = \omega_p^2 \left/  \gamma^3 \right. + k^2 c^2 \left(1 + \frac{1}{\gamma} \frac{\omega_p^2}{\omega^2} \frac{u^2}{c^2} \right), \label{CS-EM_1} \\
&& \omega^2 = \omega_p^2 \left/  \gamma \right. + k^2 c^2. \label{CS-EM_2} 
\end{eqnarray}
The corresponding wave polarizations are 
\begin{eqnarray} \label{dispersion_counter}
\textbf{E} = E \textbf{e}_x, & \mbox{ and }&  \textbf{B} = 0, \label{CS-pol-1} \\ 
\textbf{E} = E \textbf{e}_y, & \mbox{ and }&  \textbf{B} = E \frac{kc}{\omega} \textbf{e}_z, \label{CS-EM-pol_1} \\
\textbf{E} = E \textbf{e}_z, & \mbox{ and }&  \textbf{B} = -E \frac{kc}{\omega} \textbf{e}_y, \label{CS-EM-pol_2}
\end{eqnarray}
where $\textbf{e}_x$, $\textbf{e}_y$, and $\textbf{e}_z$ are unit vectors.

The key result is that all three waves have different dispersion relations. The transversely polarized electromagnetic waves described by Eqs.~(\ref{CS-EM_1}) and (\ref{CS-EM_2}) have different phase velocities and dramatically different cutoff densities:
\begin{eqnarray}
&& n_{crit}^{(y)} = \gamma^3 n_*, \label{cutoff-1} \\
&& n_{crit}^{(z)} = \gamma n_*, \label{cutoff-2}
\end{eqnarray}
where the upper index indicates the polarization of the electric field in the corresponding wave. Equation~(\ref{CS-EM_1}) is implicit, but it has a convenient form for determining the cutoff density. 

In agreement with the qualitative estimates given earlier in this section, the wave whose electric field is collinear with the counter-streaming electron motion drives electron current less efficiently. As a result, the corresponding cutoff density (\ref{cutoff-1}) is significantly higher than in the case where the electric field driving the current is orthogonal to the electron motion in the flow. The plasma with two counter-streaming flows is clearly birefringent, as the two transversely polarized waves have different cutoff densities and different phase and group velocities. 

\section{PIC simulations demonstrating birefringence in relativistically transparent plasmas}\label{Sec-MT}

The counter-streaming electron distribution that we have chosen provides an easy example to analyze, but it is extremely unstable. The distribution would quickly start to evolve after being initialized, so that the anisotropy \rc{and accompanying birefringence} would be very difficult to probe with a laser pulse.  In this context, Eqs. (\ref{cutoff-1}) and (\ref{cutoff-2}) should be viewed as an upper estimate for the degree of birefringence in a plasma whose characteristic electron relativistic factor is $\gamma$. \rc{However, one could envision setups in which the plasma is continuously pumped by laser or particle beams and thus anisotropy can be maintained for longer periods of time.}

\begin{figure}[htb]
	\centering
	\includegraphics[width=0.85\columnwidth]{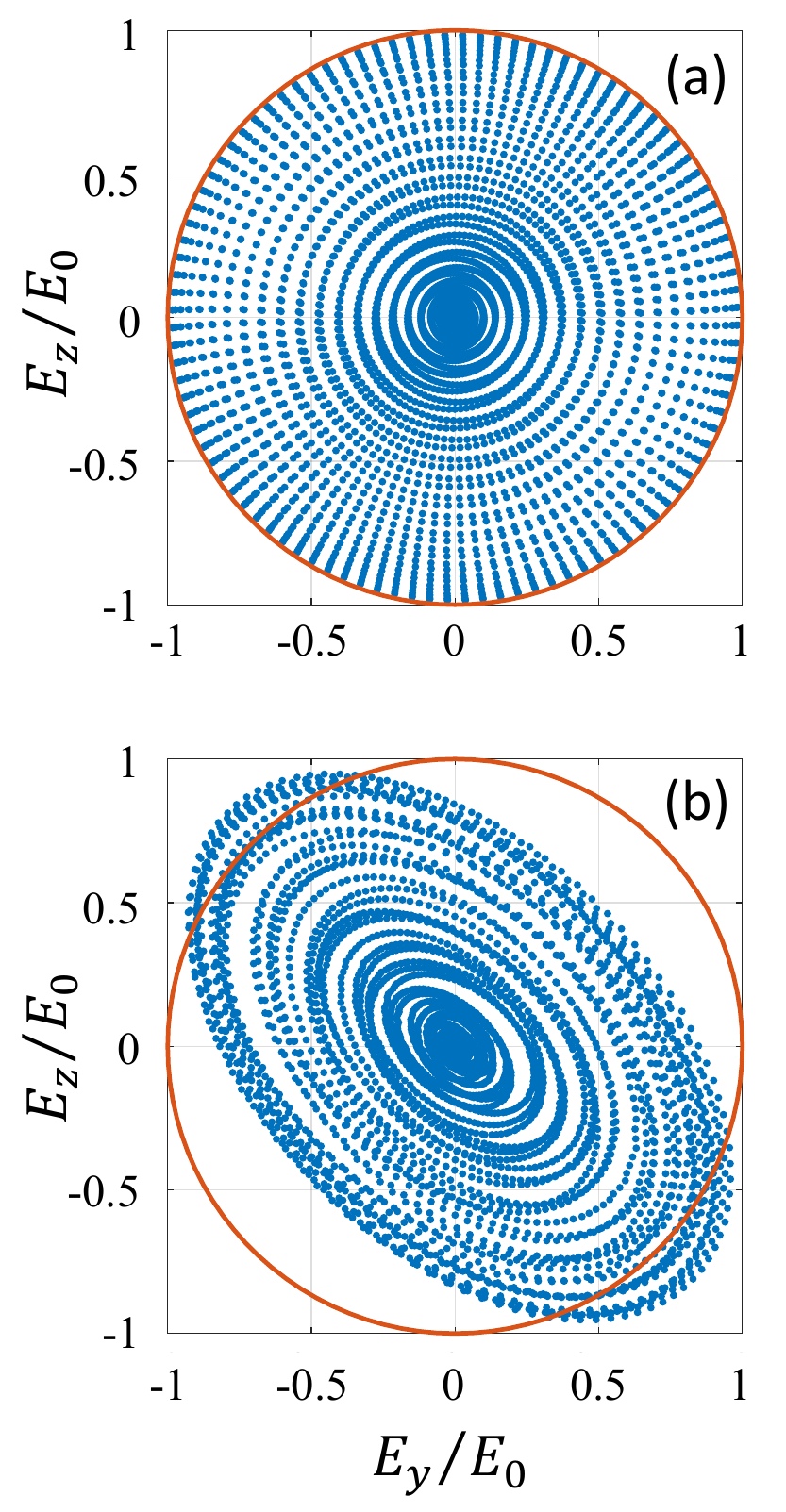} 
  \caption{Polarization of a laser pulse with an initial peak amplitude $E_0$ transmitted by plasmas with (bottom) and without (top) an anisotropy in electron momentum distribution. The solid curve shown the peak amplitude in the incoming laser pulse.} \label{FIG_RB_2}
\end{figure}

In order to further explore propagation of electromagnetic waves through a plasma with an anisotropic distribution, we have performed fully self-consistent one-dimensional PIC simulations using a fully-relativistic code EPOCH~\cite{Arber-PPCF.57.113001}. In these simulations a 50 $\mu$m thick plasma slab with $n_e = 0.4 n_*$ is irradiated by a 150 fs long circularly polarized laser pulse whose maximum electric field amplitude is $E_0 \equiv 10^{11}$ V/m. The laser wavelength is 1 $\mu$m. There are 100 cells per wavelength, each with 2000 macro-particles representing electrons and 2000 macro-particles representing ions. The ions are treated as immobile to prevent plasma expansion; this does not alter the effect of birefringence. At the same time, this ensures that the electron density remains unchanged, so that any changes of optical properties of the plasma slab are only due to changes in the electron momentum distribution. 

In the first simulation, the plasma is cold in all directions and thus the electron momentum distribution is isotropic. The polarization of the laser pulse transmitted by the plasma is shown in the upper panel of Fig.~\ref{FIG_RB_2}, where the dots represent the electric field components on the grid used by the PIC code. The maximum amplitude and the polarization of the transmitted laser pulse are the same as those of the incoming laser pulse. In the second simulation, the plasma consists of two counter-streaming electron flows aligned along the $y$-axis. The corresponding distribution is given by Eq.~(\ref{EDF_ts}), where $p_0 \approx  0.31 m_e c$  and the corresponding flow velocity is $u = 0.3 c$. This highly unstable electron distribution is given 100 fs to relax to a slowly evolving distribution before it is irradiated by the laser pulse. The polarization of the laser pulse transmitted by the plasma is shown in the lower panel of Fig.~\ref{FIG_RB_2}. In contrast to the previous case, the plasma changes the laser polarization to elliptical due to the phase velocity discrepancy between the $y$ and $z$-polarizations. The maximum amplitude of the electric field also changes, as it becomes higher than that in the incoming laser pulse (shown with a solid curve).

\rc{We can therefore conclude that the anisotropy that results from a highly unstable two-stream electron distribution produces the analytically predicted optical changes in a mixed polarization laser pulse. This is in spite of the plasma isotropizing due to the unstable plasma distribution.} The anisotropy in the simulation is sufficiently long-lived to be probed by a 150 fs long laser pulse. It takes over 300 fs for the laser pulse to pass fully through the considered plasma slab. 

It might appear that a relativistic plasma flow represents a stable case of a birefringent relativistic plasma. This is however not the case, because the flow can be eliminated by considering wave propagation in a frame of reference moving with the flow velocity. Since there is no birefringence without the flow and since all inertial frames of reference are equivalent, two electromagnetic waves with different polarizations propagating in the same direction through a plasma flow would have the same phase and group velocities. In other words, the optical properties are independent of the polarization. The same result regarding the absence of birefringence in a flowing plasma can be confirmed using a standard perturbative analysis given  in Appendix \ref{App-A}.

\begin{figure}[htb]
	\centering
	\includegraphics[width=0.99\columnwidth]{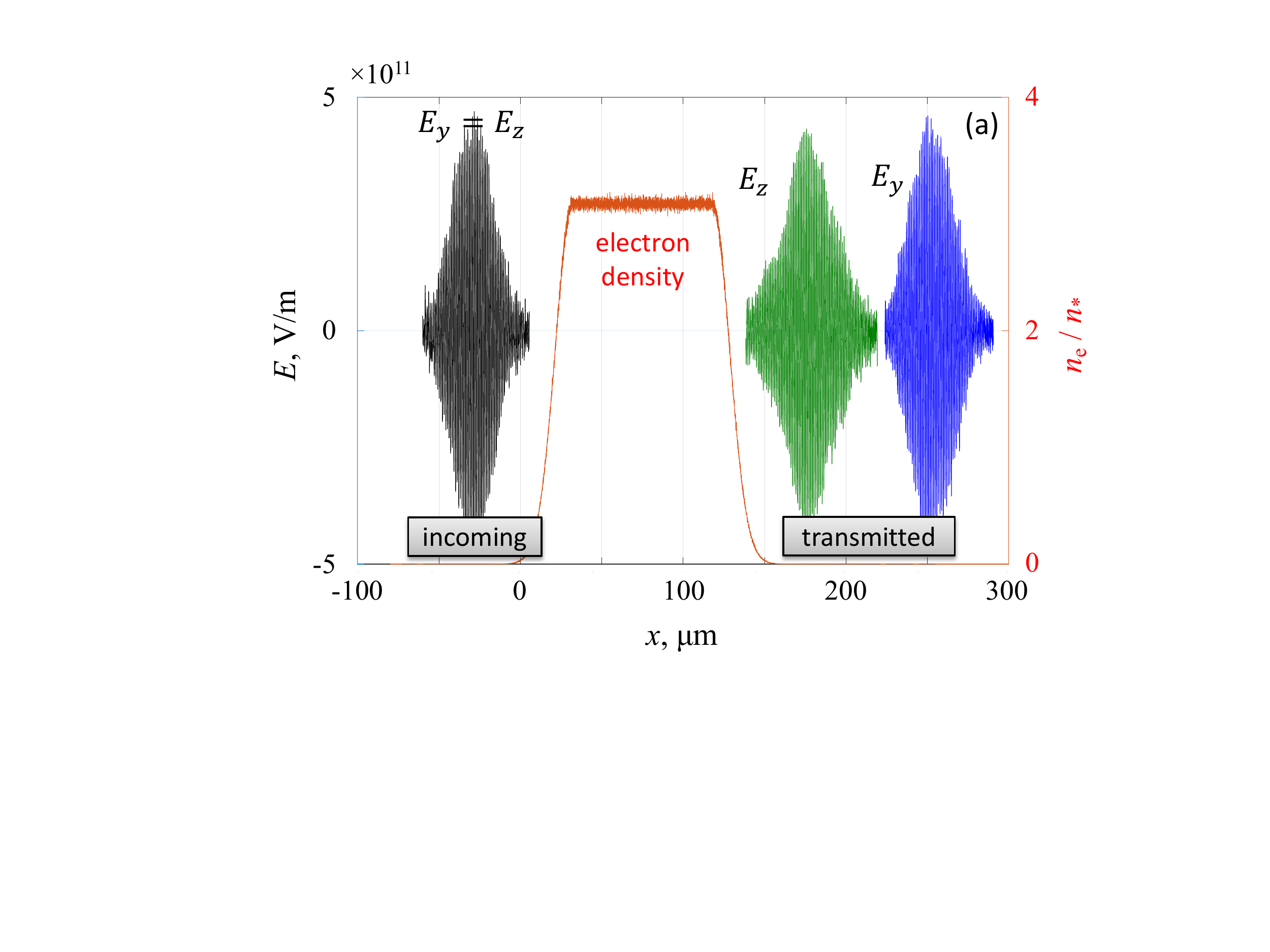}  \\
	\includegraphics[width=0.7\columnwidth]{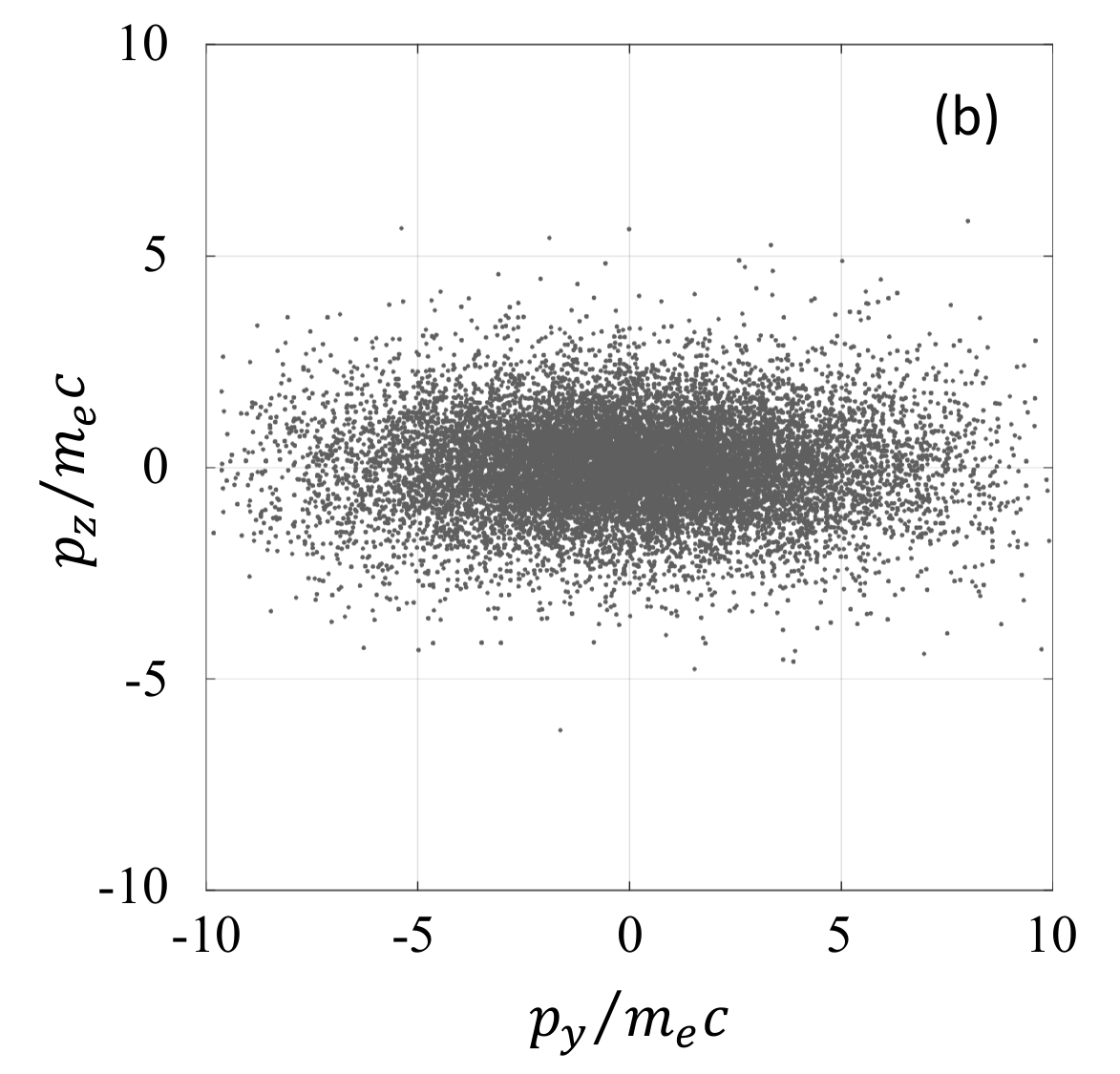}
  \caption{(a) Laser pulse splitting due to plasma birefringence in a plasma with an anisotropic electron momentum distribution in the plane transverse to the pulse propagation. \rc{(b) Initialized electron distribution function for the target plasma in ($p_y$, $p_z$) space.}} \label{FIG_split}
\end{figure}

The key conclusion then is that a relativistically induced birefringence requires an anisotropy of the irradiated electron momentum distribution that excludes an overall electron flow. However, the anisotropy is not required to be very severe in order to produce spectacular results. The final example of this section is designed to illustrate just that.

In this 1D PIC simulation, electrons are initialized with an anisotropic distribution function
\begin{equation}
    F = \frac{n}{N(\alpha,\beta)}\exp\left(-\alpha\sqrt{1+\frac{p_z^2+\beta(p_x^2+p_y^2)}{m_e^2 c^2}}  \right),
\end{equation}
where 
\begin{equation}
N(\alpha,\beta)=\int\exp\left(-\alpha\sqrt{1+\frac{p_z^2+\beta(p_x^2+p_y^2)}{m_e^2 c^2}}\right) d^3p
\end{equation}
is a normalization factor. Here $1/\alpha$ is an effective temperature normalized to $m_e c^2$ and $\beta \neq 1$ introduces anisotropy into the distribution. We set $\alpha=2.0$ and $\beta=0.1$, so that effectively the $x$ and $y$ directions are ``hotter" than the $z$ direction. We sampled this distribution 15000 times and show the scatterplot (see Fig.~\ref{FIG_split}b) of the points in $(p_y,p_z)$ space to illustrate the greater momentum spread in the $y$-direction. 

A plasma with such an initial electron momentum distribution is irradiated by a linearly polarized laser pulse propagating along the $x$-axis. The laser electric field is polarized at a 45$^{\circ}$ angle in the transverse $(y,z)$-plane, such that $\max (E_y) = \max (E_z) = E_0 = 4.5 \times 10^{11}$ V/m. The incoming pulse is 250 fs long with a wavelength of 1 $\mu$m. Note that the noise in the incoming pulse shown in Fig.~\ref{FIG_split} is physical and it is caused by the fields emitted by the plasma itself as its electron distribution evolves. The electron density profile is shown in Fig.~\ref{FIG_split}a, where the slab is 150 $\mu$m thick and the peak electron density is $3.1n_*$. There are 184 cells per micron with 200 electron and 100 ion macro-particles per cell. The ions are again treated as immobile.

A laser pulse transmitted by the plasma is shown in Fig.~\ref{FIG_split}a. Remarkably, we now have two linearly polarized laser pulses instead of just one. The leading pulse is polarized along the $y$-axis, whereas the trailing pulse is polarized along the $z$ axis. This is because the plasma is more transparent for the $y$-polarized part of the incoming pulse, so its group velocity is faster than for the $z$-polarized part of the pulse. This difference causes the two pulses to eventually split. In our previous example the propagation was not long enough for this effect to accumulate, but the corresponding phase velocity discrepancy resulted in a single elliptically polarized pulse. 


\section{Summary and Discussion}\label{Sec-C}

In this work we have examined a fundamental physical phenomena resulting from the so-called relativistically induced transparency. It is shown that the relativistic transparency impacts the laser propagation by causing the irradiated plasma to become birefringent when there is a thermally anisotropic electron momentum distribution. This birefringence can lead to such easily observable features as a change of polarization and pulse splitting, where a single pulse splits into two staggered pulses with orthogonal linear polarizations. The ability of a plasma to affect such optical changes on a pulse without damaging any equipment provides an attractive option for future optical devices.

\rc{The presented 1D PIC simulations are intended to illustrate the striking features of this birefringence. While the simulations do capture the evolving plasma distributions from the inherent instabilities, 3D simulations would provide the most robust description of this instability-driven isotropization timescale. These were performed in Ref.~\onlinecite{Stark-PhysRevLett.115.025002}, in which the anistropy was shown to persist long enough to be probed by the selected laser pulses. With this context, the 1D simulations here complement this prior work and can provide both an upper limit to the birefringence and also an example of what can occur in continuously pumped systems in which anisotropy is better maintained. }

With the recent experimental demonstration of this birefringent phenomenon~\cite{Gonzalez2016,duff2020high}, characterizing momentum anisotropy in relativistic plasmas also becomes possible. Across the spectrum of high-intensity laser-plasma interactions, many variations of plasma distribution functions are generated. These distributions largely influence the plasma applications under consideration, so proper diagnosis of said distributions would provide invaluable feedback for both modeling and experimental efforts~\cite{davies2019picosecond,davies2019investigation}.


\section*{Acknowledgments}

This research was supported by AFOSR (Grant No. FA9550-17-1-0382). Simulations were performed by EPOCH (developed under UK EPSRC Grant Nos. EP/G054950/1, EP/G055165/1, and EP/G056803/1) using HPC resources provided by TACC at the University of Texas.



\section*{References}
\bibliography{Japan_2017_paper-BIB}
%

\appendix
\section{Dispersion relation in a relativistic plasma flow} \label{App-A}

We consider a cold plasma flow of both species with
\begin{equation} \label{plasma_flow}
F = n_0 \delta( {\bf{p}} - {\bf{p}}_0),
\end{equation}
where $\textbf{p}_0 = m_e \textbf{u} / \sqrt{1 - u^2/c^2}$ is the electron momentum associated with the flow velocity \textbf{u}.  The corresponding dielectric tensor readily follows from Eq.~(\ref{disp_tensor_1}) and it is given by
\begin{eqnarray} \label{epsilon_flow}
\varepsilon_{\alpha \beta} &=& \delta_{\alpha \beta} \left( 1 - \frac{1}{\gamma} \frac{\omega_p^2}{\omega^2} \right) \nonumber \\
&+& \frac{1}{\gamma} \frac{\omega_p^2}{\omega^2} \frac{u_{\alpha} u_{\beta}}{c^2}
\frac{\omega^2 - k^2 c^2}{\left( k_{\mu} u_{\mu} - \omega \right)^2}  \nonumber \\
&+& \frac{1}{\gamma} \frac{\omega_p^2}{\omega^2} \frac{k_{\alpha} u_{\beta} + k_{\beta} u_{\alpha}}{k_{\mu} u_{\mu} - \omega}.
\end{eqnarray}

One can derive wave dispersion relations for an arbitrarily directed wave vector $\textbf{k}$ with respect to $\textbf{u}$ using the dielectric tensor given by Eq.~(\ref{epsilon_flow}). Here we consider the case where the wave propagation is transverse to the plasma flow. Let us also assume without any loss of generality that both \textbf{k} and \textbf{u} are in the $(x,y)$-plane and that $\textbf{u} = u \textbf{e}_y$ and $\textbf{k} = k \textbf{e}_x$. It follows from Eq.~(\ref{dispersion_general}) that the dispersion relations of propagating waves must satisfy the following equation:
\begin{equation} \label{dispersion_combo_1}
\left[ 1 -  \frac{1}{\gamma^3} \frac{\omega_p^2}{\omega^2} \right] 
\left[ 1 -  \frac{1}{\gamma} \frac{\omega_p^2}{\omega^2} - \frac{k^2 c^2}{\omega^2} \right]^2 = 0,
\end{equation}
where we explicitly take into account that $\gamma  = (1 - u^2/c^2)^{-1/2}$. The dispersion relations that follow from Eq.~(\ref{dispersion_combo_1}) are
\begin{eqnarray}
&& \omega^2 = \omega_p^2 \left/  \gamma^3 \right., \label{Flow_PW}\\
&& \omega^2 =  \omega_p^2 \left/  \gamma \right. + k^2 c^2. \label{Flow_EM}
\end{eqnarray}

In a reference frame moving with velocity $-u$ along the $y$-axis, the considered cold plasma flow case reduces to a well-known problem of wave propagation in a cold plasma without a flow. The three modes in such a plasma are a plasma wave ($\tilde{\omega}^2 = \tilde{\omega}^2_p$) and two electromagnetic waves ($\tilde{\omega}^2 = \tilde{\omega_p}^2 + \tilde{k}^2 c^2$), where the tilde marks quantities in the frame of reference without the flow. Using the Lorentz transformation, we find that $\tilde{\omega} = \gamma \omega$ and $\tilde{\omega}_p^2 = \omega_p^2 / \gamma$.

It is now straightforward to establish the correspondence between the waves in a plasma without a flow and the modes described by Eqs.~(\ref{Flow_PW}) and (\ref{Flow_EM}). The mode given by Eq.~(\ref{Flow_PW}) corresponds to the plasma wave. In the presence of a flow, the plasma wave also involves transverse electric field oscillations. The modes whose dispersion relation is given by Eq.~(\ref{Flow_EM}) correspond to the electromagnetic waves (note that $\tilde{\omega}^2 - \tilde{k}^2 c^2 = \omega^2 - k^2 c^2$). The mode whose electric field is polarized along the flow now also includes longitudinal electric field oscillations.

Evidently, there is no birefringence for the electromagnetic modes in a cold plasma with a flow. Both modes have the same dispersion relation, meaning that their group and phase velocities are identical.

\end{document}